\documentclass{article}

\usepackage[english]{babel}

\usepackage[letterpaper,top=2cm,bottom=2cm,left=3cm,right=3cm,marginparwidth=1.75cm]{geometry}

\usepackage{amsmath}
\usepackage{natbib}
\usepackage{graphicx}
\usepackage[colorlinks=true, allcolors=blue]{hyperref}
\usepackage{adjustbox}
\usepackage[normalem]{ulem}
\useunder{\uline}{\ul}{}

\title{Testing the Tools of Systems Neuroscience on Artificial Neural Networks}
\author{Grace W. Lindsay}

\begin{document}
\maketitle

\begin{abstract}
Neuroscientists apply a range of common analysis tools to recorded neural activity in order to glean insights into how neural circuits implement computations. Despite the fact that these tools shape the progress of the field as a whole, we have little empirical evidence that they are effective at quickly identifying the phenomena of interest. Here I argue that these tools should be explicitly tested and that artificial neural networks (ANNs) are an appropriate testing grounds for them. The recent resurgence of the use of ANNs as models of everything from perception to memory to motor control stems from a rough similarity between artificial and biological neural networks and the ability to train these networks to perform complex high-dimensional tasks. These properties, combined with the ability to perfectly observe and manipulate these systems, makes them well-suited for vetting the tools of systems and cognitive neuroscience. I provide here both a roadmap for performing this testing and a list of tools that are suitable to be tested on ANNs. Using ANNs to reflect on the extent to which these tools provide a productive understanding of neural systems---and on exactly what understanding should mean here---has the potential to expedite progress in the study of the brain.  
 
\end{abstract}

\section{Introduction}
The scale of neurophysiological and neuroimaging experiments is growing, with increasing numbers of brain regions and neurons being recorded under a variety of conditions and tasks \citep{weisenburger2019volumetric,demas2021high,van2012human,steinmetz2018challenges,bae2021functional}. This growth is undoubtedly beneficial for the quest to understand how neural structures and activity give rise to complex behavior, but it comes with important challenges. Progress has been made, for example, in making all of this data openly available and easily accessible in standardized formats and even in providing shared infrastructure for processing it \citep{abe2021neuroscience}. But the deeper question of how to best analyze these datasets remains open. 

A somewhat de-facto toolbox has arisen in systems neuroscience, with certain flavors of analysis being used across many different studies. Furthermore, novel methods aimed directly at tackling high-dimensional neural data are constantly being developed \citep{paninski2018neural}. A key question that is not frequently asked directly, however, is: are these methods helping us make progress towards better understanding of neural systems? This is, importantly, a separate question from whether these methods are technically sound (which is addressed through an evaluation of the mathematical properties of the methods and the data they are being applied to). It is also subtly separate from the question of whether these methods yield interesting results when applied to neural data. Rather, this is a question of whether applying these tools of systems neuroscience is routinely providing actionable insights into how complex high-dimensional neural systems produce interesting and adaptive behavior.  

While scientific facts are expected to be validated through repeated demonstration of their truth in multiple different experiments, scientific tools tend to go through less explicit tests of their utility. Furthermore, the same way publication bias can lead to certain false findings remaining unchallenged, not knowing the full range of analyses that have been unsuccessfully applied to a dataset may give a skewed sense of how useful certain analyses are. A more explicit research program centered on documenting in an unbiased way the extent to which different tools have yielded insights into the functioning of neural circuits would benefit the field.       

Here, I argue that artificial neural networks (ANNs) are a good testing ground for the tools of systems neuroscience. ANNs are inspired by the physiology and structure of real neural networks and can be trained to perform difficult and biologically-relevant tasks. As is detailed in Section \ref{suit}, there are good reasons to believe that tools that are capable of elucidating a satisfying understanding of ANNs may be capable of doing the same on real neural data. In opposition to real neural circuits, however, ANNs are fully observable and open to rapid experimentation, making iterating on tool testing and development with them a much faster process than relying on real neural data. To make this proposal concrete, Section \ref{whathow} details strategies for carrying out this testing including specific tools to test and how to categorize the results.     


\section{The Difficulty of the Problem}
The brain is an evolved high-dimensional, nonlinear, hierarchical, and recurrent dynamical system that produces activity varying on multiple temporal and spatial scales which gives rise to dynamic and adaptive organism-level behaviors. This makes it very difficult to understand. Some aspects of this difficulty have been formally proven \citep{rich2021hard,ramaswamy2019algorithmic}. But one only needs to attempt to make sense of a set of neural recordings collected during some interesting task to experience this difficulty first hand.

High-dimensional nonlinear systems may be hard to understand, but they are easy to find stories in. A wide range of analysis methods can be applied to a given neural data set, each returning a different take on how the underlying neural system works (see, e.g, \cite{botvinik2020variability}). We need to know if the stories our tools are helping us find are true, and not merely enticing fictions. Without well-honed tools we are at risk of compounding errors that waste years of scientific resources. Previous reflections on the state of the methods of systems biology and neuroscience give reason for concern \citep{lazebnik2002can,jonas2017could}.


Importantly, we cannot rely on informal measures of progress when assessing our tools. As philosopher Carl Craver writes: “All scientists are motivated in part by the pleasure of understanding. Unfortunately, the pleasure of understanding is often indistinguishable from the pleasure of misunderstanding. The sense of understanding is at best an unreliable indicator of the quality and depth of an explanation” \citep{craver2007explaining, thompson2021forms}. This is in part because our pre-conceived notions and expectations can bias what form of explanation we find satisfying, and blind us to more accurate yet less pleasing answers \citep{gershman2021just}.

\subsection{What are we aiming for?}
To decide on what are good methods we need to have some sense of what kind of understanding we are aiming for and a way to evaluate it. Clearly neuroscientists are aiming for something when they perform their experiments and apply certain analysis methods. What exactly that something is is rarely explicitly stated, but it usually relates to a simplified story about how the system being studied processes or transforms information. Some studies even provide cartoon diagrams or flowcharts showing how their work has updated the previously-existing picture of how a neural system contributes to the production of intelligent behavior. Such simple mechanistic stories obviously discard many details or complexities in the data, but they still provide scientists with the ability to `think through' the system and make new experimental predictions.

Insofar as what is sought by neuroscientists can be described as an abstracted set of steps that the neural system follows to achieve its computational goals, it is roughly aligned with the algorithmic level of understanding as defined by Marr \citep{marr1982vision,love2015algorithmic}. While not an algorithm in the technical sense, satisfactory algorithmic descriptions of neural datasets involve abstracting away many of the specifics of the activity patterns to provide a more compressed and logical description of the information transformations implemented by the neural circuit. 

An example of such an explanation is the description of the ventral visual stream as `untangling' representations in order to achieve invariant object recognition \citep{dicarlo2007untangling}. This way of understanding how object recognition is achieved provides a picture of the role each region of the ventral stream plays in terms of how it progressively untangles the activity of images of the same object into nicely smoothed and separated manifolds. Though it is not a quantitative model, conceptualizing the ventral stream this way still allows for experimental predictions and can be built on to incorporate other computations the ventral stream performs.  

Of course, what counts as a satisfactory algorithmic-level understanding will be subjective, but pinpointing this level as the rough goal of many studies in systems neuroscience at least offers some guiding constraints. Once we assume that goal, we can ask if our tools are providing us accurate understanding of this kind.
 

\section{Suitability of ANNs}
\label{suit}
ANNs are networks of interconnected neuron-like units. Their large-scale architecture is generally set by the experimenter. The individual weights between units are determined through a learning algorithm that makes the network as a whole capable of performing a task of interest. ANNs are not exact replicas of biological neural circuits---no model is. But they do not need to be exact replicas in order to be useful as a testing ground for our tools. Rather, the way an ANN functions simply needs to not violate the assumptions that are built into the tools. That is, as long as the data produced by an ANN has the right properties as specified by the analysis method, there is no \textit{a priori} reason we shouldn't be able to test its usefulness on ANNs. 

Importantly, this does not mean that ANNs and the brain need to work the same way even on an abstract level. If we imagine two different animals (even from two different species) performing the same cognitive task, we would expect a useful analysis tool to be able to be applied to data from both and reveal something about the mechanism(s) at play. This may reveal that both animals use the same cognitive mechanism, or it may reveal a difference between the two. In either case, the tool was still validly applied and yielded useful insights on both animals (indeed, the main way we determine if two systems are working differently is by submitting them to the same analysis). In the same way, an analysis tool may be proven useful through testing on ANNs and then when applied to real neural data reveal that the brain works differently than those ANNs.

While ANNs do not need to work just like the brain in order to be useful for testing tools, they do have several traits that make them particularly well-suited for this role, which are reviewed below.

\subsection{Full observability and perturbability}  
The study of biological neural systems is hindered in significant ways, especially historically, by the challenge of observing and perturbing neural activity. Such experimental limitations introduce multiple forms of uncertainty, put restrictions on the types of analyses that can be performed, and slow the process of building off the insights of past analyses. The activity and connectivity of ANNs, on the other hand, can be perfectly observed simultaneously across all units in the network under any experimental condition. A large variety of manipulations are easily implementable including connectivity changes, activity perturbations, and even `developmental' interventions during network training. The situation with ANNs thus far exceeds our experimental abilities for even the simplest of organisms, such as \textit{C. elegans}.  

\begin{figure}
\centering
\includegraphics[width=0.75\textwidth]{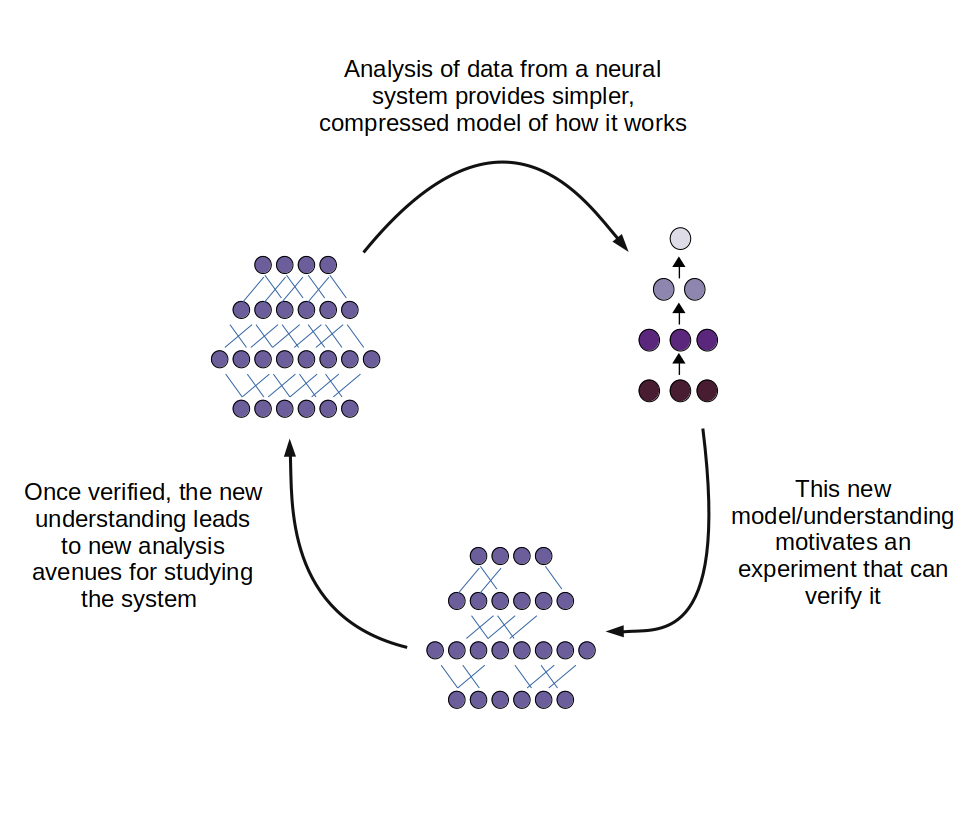}
\caption{\label{closedloop} Experimentally-validated understanding. The utility of an analysis method can be evaluated by how well it contributes to this process of understanding. }
\end{figure}

The removal of experimental barriers expands the number and type of analyses that can be performed, but more importantly it makes a process of 'experimentally-validated' understanding possible. Here, experimentally-validated understanding refers to verifying the insights gleaned from an analysis method through further experimentation based on those insights (Figure \ref{closedloop}). Specifically, assuming we believe that applying the tools of systems neuroscience should provide us with a hypothesis of how a neural system works, then we should be able to devise experiments that directly test that hypothesis. If the hypothesis turns out to be correct, then the tools were successful at providing accurate insights. If not, then the tools were successful in creating a possible narrative, but not an accurate one. While in theory this form of experimental validation of tools can occur in real neural networks, performing the exact experiment that would best test the insights provided by a given method is often infeasible, and certainly cannot be done as thoroughly and quickly as it can in artificial neural networks. 



In addition to the practical benefits, full observability and perturbability also offer an important constraint: when we have perfect data collection and perfect ability to manipulate the system we cannot default to uncontrollable external factors to fill in the holes of our explanations. It is not uncommon for studies to make references to things like neuromodulation, inputs from an unrecorded brain region, downstream readout effects, or changes in behavioral state to explain what appears unexplainable in a given dataset. With the use of ANNs, experimenters must face the fact that even with full knowledge of the low-level details of a system, we can still struggle to identify a satisfying mechanistic explanation. In other words, the exploration of ANNs can teach us a lesson we would've learned soon enough through advances in experimental methods: the recording technology was never the only bottleneck; our tools of analysis may be lacking too.

\subsection{Proven similarity to real neural networks}
The recent resurgence of artificial neural network models in computational neuroscience has stemmed in large part from their success in matching important features of real neural activity. This was shown most prominently in primate visual cortex where convolutional neural networks (CNNs) trained to perform object recognition were able to predict neural activity and match representational patterns as recorded through electrophysiology and neuroimaging \citep{yamins2014performance, khaligh2014deep}. This same approach has been fruitfully applied to explaining neural response properties in many other brain areas \citep{kell2018task,banino2018vector,nayebi2021explaining,michaels2020goal,wang2021evolving,Barak2017}.

Even before training, many of the basic properties of ANNs match those of the brain---albeit at an abstract level. Given that artificial neurons were inspired by the physiology of real neurons \citep{mcculloch1943logical}, they perform they same basic function of taking a weighted sum of inputs from other neurons and non-linearly transforming that into an output activation value. ANNs also implement parallel and distributed (and increasingly, recurrent) processing of information, as used by the cortex. And in some cases the architectural details of ANNs are directly inspired by the brain. The convolutions and pooling layers in CNNs, for example, are based on the simple and complex cells identified in cat primary visual cortex \citep{lindsay2021convolutional}.   

These basic similarities (and the fact that more biological details can be added; see Section \ref{assumps}) make it likely that many of the tools applied to real neural data will be applicable to ANNs as well. For this reason, ANNs differ from other proposed test beds of tools, such as a simulated microprocessor \citep{jonas2017could}. While, as argued above, ANNs do not need to function just like the brain in order to be amenable to the same tools, using models that have many of the same basic properties will likely expedite the process of finding the most valuable tools.  

\subsection{Uncertainty in how they work on an abstract level}
ANNs are sometimes criticized as models of the brain because they `replace one black box with another'. This critique rests on the fact that, because the weights in ANNs are trained rather than hand-designed, the principles by which these networks work remain mysterious, even when they prove capable of predicting neural activity. Knowing the distributed activity of thousands (or more) neural units does not give neuroscientists a satisfying explanation of how the system achieves the useful computations that make it capable of carrying out difficult tasks. What many neuroscientists are implicitly aiming for is thus something closer to the algorithmic level of understanding as described above. Such abstract explanations provide a compressed and human-interpretable set of steps or principles by which a network transforms its inputs into outputs.

Our ignorance of the abstract algorithms implemented by a particular brain region under a given task mirrors our ignorance of the abstract algorithms implemented by an ANN. It may seem that this makes ANNs a poor model to test the methods by which we identify such algorithms. Indeed, a standard and important part of methods development is testing on data where 'ground truth' is known. This is usually done by creating synthetic data with certain properties and showing that the method can recover those properties, or using real data whose ground truth have been well-established through years of research.

I argue that being in the same state of ignorance toward a trained ANN as we are toward the brain is a benefit of these models. This is because knowing `the correct answer' in advance can lead us to falsely conclude that a method has led us to that answer, when in fact we were guided at least in part by our pre-existing knowledge. A focus on clean, well-behaved synthetic data also takes certain types of answers off the table. Specifically, by using hand-selected and digestable test data we run the risk of designing methods that find the kind of simple mechanistic descriptions we already expect to find, when in fact the full space of algorithms implementable by artificial or biological neural networks is vast and largely unexplored. Proving that a method can reveal a mechanism that we had not expected, nor even thought of, is a much stronger demonstration of its power than showing it can reveal answers we already knew. 

With the lack of ground truth, however, how do we verify that the methods were successful? This is where the importance of experimentally-validated understanding as described above comes into play. As long as a method produces an understanding of neural network function that is proven correct through further experimentation, we can feel confident in its utility.  

Another important outcome that could arise from a concerted effort to test how our tools lead to understanding is a more explicit definition of understanding itself \citep{craver2007explaining,stinson2017mechanistic}. Because we don't know the form the answer will take prior to analyzing an ANN, we will need to decide when we have achieved a satisfactory answer. As mentioned above, what systems neuroscientists are aiming for is rarely explicitly discussed. Such reflection on what really counts as understanding could help clarify the goals of systems neuroscience and orient our tools around those goals. This is especially important as some scientists have questioned the amount of progress neuroscience has made in recent decades; knowing what our goals our makes it much clearer to determine if we are progressing towards them. By using ANNs in this way we can also benefit from recent---and not so recent (e,g, \cite{Andrews1995})--- work stemming from the machine learning literature, which has also grappled with this very question of what it means to understand a neural system \citep{thompson2021forms, leavitt2020towards,lillicrap2019does, olah2018building,cao2021explanatory,cao2021explanatory2}.

It is sometimes claimed that we should not expect to get a satisfying human-comprehensible understanding of ANNs. According to this view, the networks are simply too distributed and unconstrained to promise a simpler description of their function. If this goal is not achievable through any means then it of course will not be achievable by using the tools of systems neuroscience. If we are to assume that ANNs can't be understood, however, that raises the question of what makes the brain any different. Is the goal of a compressed understanding of the mechanisms of computation in the brain as ill-fated as such an understanding of ANNs? Some scientists would say yes (from \cite{lillicrap2019does}: `a lot of approaches currently deployed
to understand brains may be rather transparently unable to deliver the results neuroscientists are looking for.') and that therefore neuroscientists should largely abandon this goal and instead aim to describe the architectures, objective functions, and learning rules that give rise to these systems \citep{richards2019deep}. On the other hand, there may be differences between ANNs and the brain that bear on this question (discussed in Section \ref{assumps}). I argue that while there is no guarantee that a satisfying description of the function of any trained network is necessarily possible, it seems reasonable to assume that we should at least be able to find some description of how a neural network works that is more compressed than simply listing all of its weights and activity values---and that this is true of the brain as well. What's more, if extensive analysis of ANNs using the tools of systems neuroscience does prove that simplified understandings are largely out of reach, this should be considered a relevant finding for neuroscientists to reflect on.

\subsection{Previous Insights from ANNs}
In addition to the above arguments for the suitability of ANNs as tests of the tools of systems neuroscience, we can look to cases in which the study of ANNs has already shed light on the ways we understand the brain.

One beloved tool of systems neuroscience that has been critically-examined in ANNs is single-cell selectivity. For decades, the strength and quality of the tuning of individual neurons has been assumed to be important for understanding a brain region's function. A variety of studies investigating the relationship between the quality of single-cell tuning and task performance in ANNs have shown these single-cell properties to be of far less importance than their emphasis in the neuroscience literature would suggest---and that in fact strong single-cell selectivity can actually have a negative relationship with performance \citep{morcos2018insights,morcos2018importance, amjad2021understanding, Lindsay2018, leavitt2020selectivity}. This shows how historical experimental limitations (e.g. the ability to only record from single neurons at a time) can counter-productively constrain the tools of today. It also demonstrates that basic intuitions (e.g. that a cell responding strongly to a stimulus indicates its functional importance for processing that stimulus) can't always be trusted when studying complex nonlinear systems.      

A recent study on recurrent ANNs explored the properties of representational similarity analyses as a means of comparing neural systems \citep{maheswaranathan2019universality}. This work showed that aspects of representational geometry are influenced by features of the networks that are not related to its function, calling into question the usefulness of these tools for developing productive mechanistic understanding. On the other hand, they found that an analysis of certain features of the dynamics did provide more consistent insights into the computations being performed. 


Additionally, previous works that have developed and reflected on specific analysis tools have indeed acknowledged the usefulness of testing these tools on ANNs \citep{Bernardi2020,zaharia2021visualizing,Barrett2019,schaeffer2020reverse}. According to \cite{chung2021}, for example, `ANNs can serve as a testbed for developing population-level analysis techniques, such as geometric approaches, even if they are ultimately aimed at neuroscience applications.' This idea is therefore intuitive to some scientists working to understand both real and artificial neural systems. In this piece, I have provided a full account of the justification behind this intuition, and will now provide a more detailed roadmap of how to carry this idea out. 



\section{What Tools to Test and How}
\label{whathow}
Not all analysis methods applied to neural data will be applicable to ANNs, and therefore not all can be vetted on ANNs. Understanding the properties and aims of an analysis method is a crucial first step in vetting it. Once a method is selected, we then want to know the circumstances under which it provides productive results. For this it is important to test a tool on many different ANNs with different architectures and trained on a range of tasks.

\begin{figure}
\centering
\includegraphics[width=0.8\textwidth]{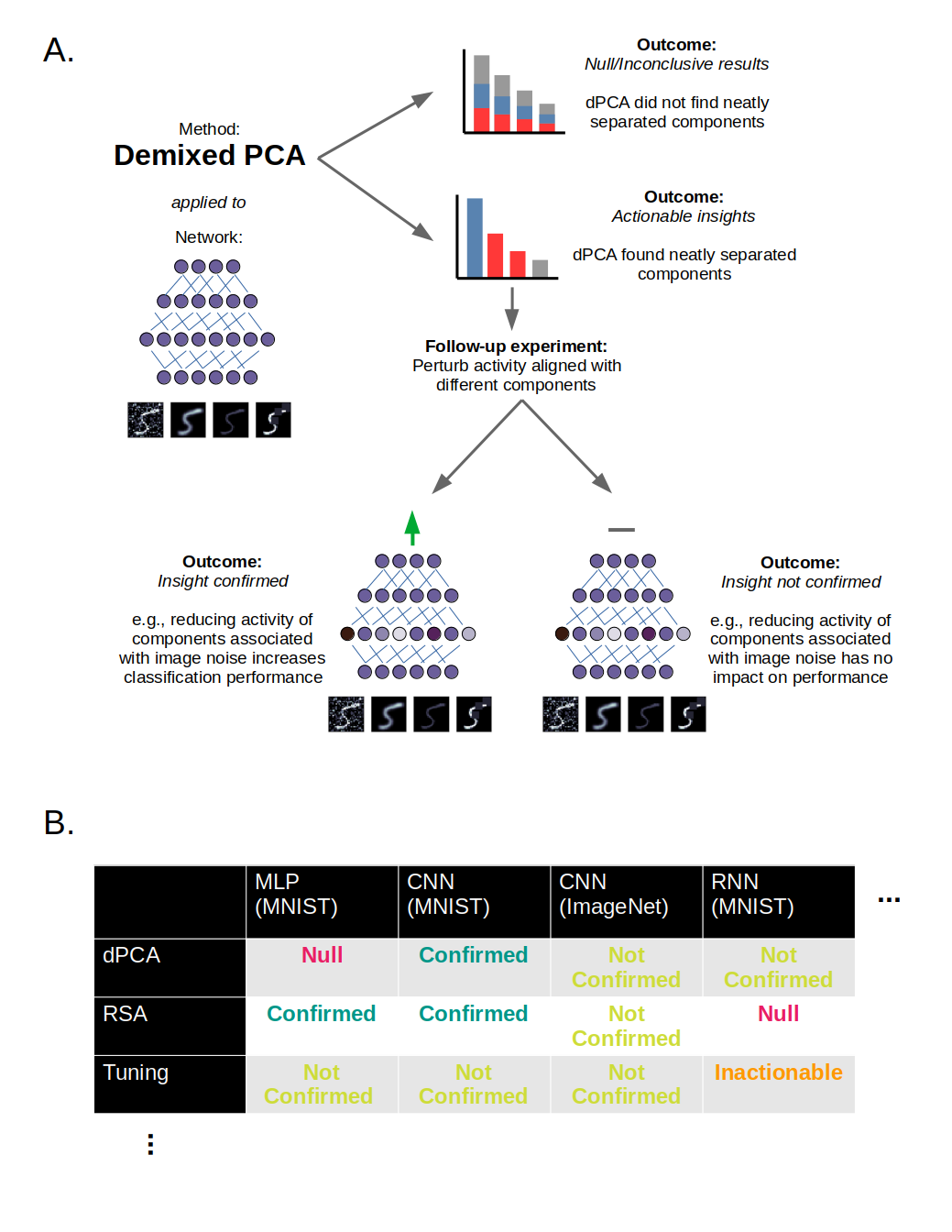}
\caption{\label{outcomes} A. Example of how to test a given method on a given network. Here the method is demixed PCA and the network is a feedforward network trained to classify images of digits while ignoring different types of noise added to them. dPCA provides components whose variance can be associated with different task parameters (blue = digit identity, red = image noise type, gray = interaction). If all the dPCs remain mixed (top bar graph) then the analysis outcome is inconclusive. If they are well demixed (bottom bar graph) then the analysis has provided actionable insights. These can be acted on by, for example, perturbing activity based on these components. If these perturbations have the expected impact on classification performance (left fork) then the analysis method has been successfully vetted according to the standard of experimentally-validated understanding. B. Results of testing many tools on many networks (MLP, CNN, and RNN are network architectures; MNIST and ImageNet different tasks). A given study may want to focus on how well a single tool fairs against many different networks (e.g. a row in this table) or on which tools do best on a specific network type (column). }
\end{figure}

Such a research program would document the utility of a given tool for understanding different types of neural circuits (Figure \ref{outcomes}B, row-wise). An alternative desire could be to determine which tools are best for understanding a specific type of circuit. In this case, a single ANN would be submitted to a range of analysis tools and the outcomes of each would be documented (Figure \ref{outcomes}B, column-wise). 

In either case, the outcome associated with the application of a given method is determined by its ability to provide experimentally-validated understanding as described above. Specific details of how this process would proceed are below.

\subsection{Checking assumptions}
\label{assumps}
For most methods, there are simple properties the data must have in order for that method to be validly applied to that data. To determine if ANNs can be validly be submitted to an analysis method we therefore need to know if they violate any of the method's assumptions or requirements. For this, it is good to review the basic working properties of ANNs.

ANNs are parallel distributed processing systems. As described above, an individual artificial neuron takes in a weighted sum of inputs from other neurons (which can be thought of as akin to subthreshold membrane potential) and produces an output via a nonlinear function of that input (and possibly other factors). The activity of artificial neurons is usually non-negative and continuous and can therefore be thought of as representing a firing rate (though spiking ANNs and those that allow negative activity values do exist). The networks tend to be modular, at least in the sense that neurons are separated into layers with constraints on their connectivity defined by those layers. Purely feedforward networks have no inherent temporal dynamics, though simple cell-intrinsic dynamics can be added. Recurrent networks (wherein connections are allowed to exist within a layer or from a later layer back to an earlier one) do have dynamics, which are usually temporally discrete. It is also possible to study the dynamics of learning in these networks, by recording data at various points in the training procedure. Most ANNs do not have any form of intrinsic noise, leading to deterministic responses to input stimuli. Weights generally do not obey Dale's law---that is, individual artificial neurons can produce both positively and negatively-weighted connections.

\begin{table}[ht]
    \centering
    \begin{tabular}{p{0.95\linewidth} }
      \textbf{The Toolbox}  \\ \hline
      The following are (not mutually-exclusive) classes of tools commonly-applied to neural data that would be ammenable to testing on artificial neural networks:  \\
      \textit{\textbf{Dimensionality Reduction.}} Many different flavors of dimensionality reduction have been developed with applications to neuroscience. Dimensionality reduction is useful for visualizing high-dimensional data, discerning properties of the representation, and eliminating noise. See: \cite{williamson2019bridging,jazayeri2021interpreting,pang2016dimensionality,cunningham2014dimensionality,Goddard2018,Gao214262}  \\
      \textit{\textbf{Latent Factor Modeling.}} Identification of latent factors has some overlap with dimensionality reduction techniques in that it aims to find a small set of factors that explain much of the data. However these methods are usually based on probabilistic models, incorporate latent dynamics, and allow for more nonlinear relationships between the latent factors and activity. See: \cite{paninski2018neural,linderman2017using, Whiteway2019, hurwitz2021building,Recanatesi2021} 
       \\
      \textit{\textbf{Representational Similarity Analyses.}} While these methods have been popular as a means of comparing ANNs to neural activity, they can also be applied to look for theoretically-motivated encoding schemes or compare across different neural populations within a single system, thereby providing insights into how information is transformed. See: \cite{Kriegeskorte2008,kornblith2019similarity,morcos2018insights,Kriegeskorte2013,kriegeskorte2021neural}
       \\
      \textit{\textbf{Representation Geometry}} A diverse set of analyses that characterize the geometry of neural population responses are becoming more common as a means of understanding the computations and transformations the brain is implementing. See: \cite{Bernardi2020,chung2021,nieh2021geometry,chaudhuri2019intrinsic}
      \\
      \textit{\textbf{Network Analyses.}} Network analyses are influenced by methods from graph theory and can be applied on structurally or functionally defined networks in order to reveal non-trivial features of their topology. These methods would benefit most from testing on recurrent artificial neural networks. See: \cite{Bassett2017,Medaglia2015, Bassett2018,Sporns2014}
      \\
      \textit{\textbf{Encoding Models.}} Methods for quantifying the type and amount of information encoded in a given population's activity can include characterization of tuning properties, trained decoder performance, regression models, and formal metrics of information theory. See: \citep{kriegeskorte2021neural,Kriegeskorte2019enc,Butts2006,QuianQuiroga2009,Timme2018,paninski2007statistical}
      \\
      \textit{\textbf{Bespoke Methods.}} Many experimental studies have analyses designed specifically for the data collected in the study, and these methods normally don't undergo formal method development. ANNs built to replicate the experimental setup could be a good means of exploring the utility of these methods.    
    \end{tabular}
    \caption{\label{toolbox_tab} A toolbox of systems and cognitive neuroscience}
\end{table}

These basic features make ANNs immediately subjectable to certain common analyses in systems neuroscience. For example, visualization of activity via dimensionality reduction techniques such as PCA has been used to gain intuition about the response properties of layers with hundreds or thousands of units---the same way it has been used to visualize activity of real neural populations. In fact, I argue that many of the same tools applied in systems neuroscience are applicable to ANNs exactly because many systems neuroscientists operate under an ANN-like view of the brain. That is, they mainly think of a neural population as a collection of simple input-output devices operating in parallel whose computations are enacted by the activity of these units, and that activity is a result of the connections the units make amongst themselves. For more examples of popular styles of analysis that would be well-suited to vetting through application to ANNs, see `the Toolbox' (\ref{toolbox_tab}).  

Not all common forms of analyses are immediately applicable to ANNs however. Studying noise correlations in these networks, for example, would be difficult: given that noise correlations rely on the existence of trialwise noise across the neural population, decisions would need to be made about how to add that noise to ANNs that don't normally have it. The specifics of these decisions would likely have large impacts on the analysis outcomes and the interpretation of the results would be generally muddied by the fact that the noise is known to not be core to the network's functioning. Studies of oscillations and analyses of local field potentials would similarly be difficult to properly replicate in ANNs.    

It is important to note that while the above basic features are true of most ANNs, many of them are technically malleable. In addition to the caveats already given, there are ways, for example, to train networks that have different neuronal subtypes and obey Dale's law \citep{Song2016,Cornford2020.11.02.364968}. The sparsity of neural responses---a feature sometimes assumed when studying real neural populations---can also be controlled through regularization. In general, if a given analysis method assumes a feature not present by default in ANNs, but one that can be thoughtfully added, then ANNs with that feature could still be a useful testing ground for the tool. 

Related to the previously-mentioned concern that ANNs simply won't be understandable in the way we desire, there may also be a concern that ANNs won't be understood using the same tools as those we apply to the brain. However, if ANNs do not violate any explicit assumption or requirement of the method then why should these methods not illuminate the workings of ANNs? Usually the answer is that there is some looser, unstated assumption about why the tool is useful for understanding the brain, and ANNs somehow violate these less formal assumptions. One such example is the fact that the brain is the result of evolution and a set of developmental processes based mainly on local learning rules; ANNs on the other hand are typically trained from random initialization with gradient descent. The question of whether these differences actually invalidate the use of the tools of neuroscience on ANNs is an empirical one. A variety of more biologically-plausible forms of training ANNs are constantly being developed \citep{Lillicrap2020bp}; these networks can be used to determine if local learning rules really do impact the applicability of our tools. In any case, it would be beneficial to make these unstated assumptions about our tools explicit, so that their application to real neural data is better understood.

\subsection{Testing procedure}
A simple open and iterative procedure will help identify the most promising tools for different questions. As mentioned above, this process can focus on identifying the best use cases for a given tool, or identifying the best tool to understand a given type of neural circuit. 

For each combination of method and network, a clear documentation of the outcome will need to be provided. This requires a method of determining different types of `success'. Elaborating on the experimentally-validated form of understanding described above, I advocate for a graded measure of the success of the outcome. 

The first and lowest rung would be to conclude that the analysis provided null or uninterpretable  results. The next possible outcome is that the analysis results are interpretable and offer some intuition, though are not precise or clear enough to lead to any actionable insights. The next level up from that would be analysis results that lead to clear actionable insights, meaning they are directly useful in guiding the design of new experiments. The final and highest tier is reserved for analysis outcomes that both guide the design of experiments and those experiments, once performed, confirm the insight initially provided by the analysis.

An example of this procedure is shown in Figure \ref{outcomes}A. Here the network in question has been trained to classify images of hand-written digits with different types of noise added. The analysis method is demixed PCA (dPCA; \cite{kobak2016demixed}), which aims to reduce population activity to a small number of components that are interpretable with respect to task parameters. When applying dPCA to activity at a given network layer, we get a set of principal components ranked according to variance explained, with a given proportion of variance within each component attributed to each task parameter. Task parameters here would be the identity of the digit in the image and the type of noise applied to it (and an interaction term). 

In Figure \ref{outcomes}, multiple potential outcomes of applying dPCA to this network are shown. If dPCA fails to demix the variance in network activity, then the analysis results are inconclusive. If well-demixed dPCs are found then the analysis has provided some intuition about how this network classifies noisy images (though not shown here, this may be especially true if dPCA reveals an interesting pattern in how the demixing is done across layers). Specifically, it supports a story of how information related to digit identity and noise type are separated into different activity subspaces. At this point, however, this remains an observational finding. To test if dPCA provides experimentally-validated understanding, it needs to provide insights that can be successfully acted on in further experiments. In this case, that could mean perturbing the activity of the network at a given layer according to the activity patterns identified by dPCA. For example, activity could be reconstructed using all but the dPCs associated with image noise. If these activity patterns are indeed associated with representing image noise and thus not important for digit identity classification, we would expect digit classification performance to increase under this perturbation. If that outcome is found, then the insights provided by dPCA are confirmed. 

Such a process is of course not perfect. There are many decisions the experimenter needs to make and subjective judgements are required, as is always the case in the scientific process. For this reason, transparency is important. The `file drawer problem' that makes us oblivious to all the times an analysis was tried and failed to produce insights must be explicitly avoided here and the reasons for all of the conclusions drawn should be clearly laid out. To facilitate this, the set of analyses and networks that will be tested should be decided in advance, and the results of all planned tests should be reported regardless of outcome. Pre-registration of studies may be an appropriate means of ensuring transparency here. 

\section{Conclusion}
The tools we use to analyze neural circuits will determine how accurately and efficiently we develop useful theories of brain function. For this reason, it is important to explicitly reflect on and vet these tools. Here I have argued that ANNs are the appropriate systems on which to perform this vetting. The application of common tools of systems neuroscience to ANNs, followed by explicit testing of the insights they provide, will determine their utility in achieving the goals of the field. What's more, this will encourage more explicit definitions of these goals, including potentially acknowledging that some goals may be out of reach.  While the focus here has been on testing existing tools, this procedure could also contribute to the development of new methods based on identified gaps (methods possibly inspired by current tools used in the machine learning literature to understand ANNs).   

\section{Acknowledgements}
Thanks to Rosa Cao, Matteo Colombo, and Josh Merel for input on the manuscript.



\bibliographystyle{alpha}
\bibliography{sample}

\end{document}